\documentclass[proceedings]{JHEP3}

\PrHEP{PrHEP hep2001}                    
\conference{International Europhysics Conference on HEP}

\renewcommand{\a}{\alpha}

\newcommand{\eps}{\varepsilon}

\newcommand{\as}{\alpha_{s}}
\newcommand{\bea}{\begin{eqnarray}}
\newcommand{\eea}{\end{eqnarray}}
\newcommand{\beq}{\begin{equation}}
\newcommand{\eeq}{\end{equation}}
\newcommand{\nn}{\nonumber}

\newcommand{\pl}{Phys.~Lett. }

\renewcommand{\prl}{Phys.~Rev.~Lett. }
\newcommand{\cO}{{\cal O}}

\newcommand{\cA}{{\cal A}}

\newcommand{\ba}{\begin{array}{c}}
\newcommand{\bat}{\begin{array}{cc}}
\newcommand{\ea}{\end{array}}
\def\eqn#1{(\ref{#1})}
\def\slashchar#1{\setbox0=\hbox{$#1$}\dimen0=\wd0%
\setbox1=\hbox{/}\dimen1=\wd1%
\ifdim\dimen0>\dimen1%
\rlap{\hbox to
\dimen0{\hfil/\hfil}}#1\else                                     
\rlap{\hbox to \dimen1{\hfil$#1$\hfil}}/\fi}
\title{  $\eps '  /\eps$ in the Standard Model}
\author{\speaker{Ignazio Scimemi}\\
Departament de F\'{\i}sica Te\`orica, IFIC,
CSIC --- Universitat de Valencia, Edifici d'Instituts de Paterna,
Apt. Correus 22085, E--46071, Val\`encia,  Spain\footnote{On leave to
  ITP--University of Bern, Sidlerstr. 5, 3012 Bern, Switzerland.}}
\author{Elisabetta Pallante\\
SISSA,Via Beirut 2--4, 34013 Trieste, Italy}
\author{Antonio Pich\\
Departament de F\'{\i}sica Te\`orica, IFIC,
CSIC --- Universitat de Valencia, Edifici d'Instituts de Paterna,
Apt. Correus 22085, E--46071, Val\`encia,  Spain}

\abstract{
We overview the detailed analysis of $\eps'/\eps$ within the
Standard Model, presented in ref.~\cite{PPS:01}.
When all sources of large logarithms are considered, both at short and long
distances, it is possible to perform a reliable Standard Model estimate of
$\eps'/\eps$. The strong S--wave rescattering of  the final pions
has  an  important impact on this observable~\cite{PPS:01,PP:00}.
The Standard Model prediction is found to be~\cite{PPS:01}
$  { \rm Re}\left(\eps'/\eps\right) =(1.7 \pm 0.9) \cdot 10^{-3}$,
in good agreement with the most recent experimental measurements.
A better   estimate of the strange quark mass would reduce
the uncertainty to about 30$\%$.
}

\begin{document}

\section{Introduction}
\label{sec:introd}
In recent times the determination of Re$\,\left(\eps '/\eps\right)$ 
has stimulated a lot of
 work both  on the theoretical and experimental sides.
The latest has been recently clarified by the new NA48~\cite{na48},
Re$\,\left(\eps'/\eps\right)=(15.3\pm 2.6)\cdot 10^{-4}$, and the KTEV~\cite{ktev},
Re$\,\left(\eps '/\eps\right)=(20.7\pm 2.8)\cdot 10^{-4}$, results. 
The present experimental world
 average is~\cite{na48}--\cite{E731}
\beq
{\rm Re}\left(\eps '/\eps\right) = (17.2\pm 1.8)\cdot 10^{-4}\ .
\label{eq:exp}
\eeq
The theoretical prediction  has been  the subject of many debates since
 different groups, using different methods or approximations  obtained
 different results~\cite{munich}--\cite{dubna}.
Recently however it has been observed~\cite{PPS:01} that once all essential ingredients are
 taken into account, including final state interactions (FSI)~\cite{PP:00}, one 
can give a reliable estimate of Re$\,\left(\eps '/\eps\right)$ 
 which is in perfect agreement with the
 experimental value. In ref.~\cite{PPS:01} a detailed analysis  is presented,
 which includes the evaluation of all large logarithmic corrections both at short and long
 distances; the resulting Standard Model prediction is
\beq
{\rm Re} \left(\eps '/\eps\right) = (17\pm 9)\cdot 10^{-4}\ .
\label{eq:SMpred}
\eeq
The subject of this  talk is a review  of the main ingredients  in the calculation
of $\eps '/\eps$.

The physical origin of $\eps '/\eps$  is at the electroweak scale, where  the
flavor--changing processes can be described in terms of quarks, leptons and
gauge bosons  with the usual gauge coupling perturbative expansion.
At the scale  $M_Z$ the heavy gauge bosons $W^{\pm}$ and $Z$, and the top quark are
integrated out of the theory.
The dynamics is then described in terms of Wilson coefficients $C_i(\mu)$ and
operators $Q_i(\mu)$, via a Lagrangian of the form
\beq\label{eq:Leff}
 {\cal L}_{\mathrm eff}^{\Delta S=1}= - \frac{G_F}{\sqrt{2}}
 V_{ud}^{\phantom{*}}\,V^*_{us}\,  \sum_i  C_i(\mu) \; Q_i (\mu).
 \label{eq:lag}
\eeq
The values of the coefficients $C_i$ are matched with the underlying theory at the  electroweak scale $\sim M_Z$.
Then, using the Operator Product Expansion (OPE)~\cite{WI:69} and renormalization group
equations~\cite{RGroup},  one can evaluate the Wilson coefficients at any
scale $\mu$ summing up the short--distance logarithms.
The overall renormalization scale $\mu$ separates
the short-- ($M>\mu$) and long-- ($m<\mu$) distance contributions,
which are contained in $C_i(\mu)$ and $Q_i$, respectively.
The physical amplitudes are independent of $\mu$; thus, the
explicit scale (and scheme) dependence of the Wilson coefficients should
cancel exactly with the corresponding dependence of the $Q_i$
matrix elements between on-shell states.

Our knowledge of $\Delta S=1$ transitions has improved qualitatively
in recent years, thanks to the completion of the next-to-leading
logarithmic order calculation of the Wilson coefficients
\cite{buras1,ciuc1}.
All gluonic corrections of $\cO(\as^n t^n)$ and $\cO(\as^{n+1} t^n)$, where
$t\equiv \log M/m$ and $M$ and $m$ are any scales appearing in the evolution, 
are already known. Moreover the full $m_t/M_W$ dependence (to first order
in $\a_s$ and $\a$) has been taken into account at the electroweak scale.
We will fully use this information up to scales $\mu\sim \cO(1\; {\rm GeV})$,
without making any unnecessary expansion.
 At a scale $\mu < m_c$  one has a three--flavor theory  described by  a
 Lagrangian  of the same general form as in eq.~(\ref{eq:Leff}).
The difficult and still unsolved problem   resides in the calculation of the
hadronic matrix elements.
As we  will see in the following the large--$N_c$ expansion and Chiral
Perturbation Theory ($\chi$PT) allow to estimate  those matrix elements
 with sufficient accuracy for the determination of $\eps'/\eps$.
In the following 
we  adopt the usual isospin decomposition:
\beq\label{eq:AI}
\begin{array}{lr}
A[K^0\to \pi^+\pi^-] \equiv  \cA_0 + {1\over \sqrt{2}}\,  \cA_2
\, , &\quad
A[K^0\to \pi^0\pi^0] \equiv  \cA_0 - \sqrt{2}\,  \cA_2\, .
\end{array}
\eeq
The complete amplitudes
${\cal A}_I \equiv A_I\, \exp{\left\{i\delta_0^I\right\}}$
include the strong phase shifts $\delta_0^I$.
The S--wave $\pi$-$\pi$ scattering generates a large
phase-shift difference between the $I=0$ and $I=2$ partial
waves \cite{GM:91}:
%
$\left(\delta_0^0 - \delta_0^2\right)(M_K^2) = 45^\circ\pm 6^\circ
\, .$
%
There is a corresponding dispersive FSI effect in 
the moduli of the isospin amplitudes, because the real and 
imaginary parts are related by analyticity and unitarity.
The presence of such a large phase-shift difference clearly signals
an important FSI contribution to $A_I$. 
In terms of the $K\to\pi\pi$ isospin amplitudes,
\beq
{\varepsilon^\prime\over\varepsilon} =
\; e^{i\Phi}\; {\omega\over \sqrt{2}\vert\eps\vert}\;\left[
{\mbox{Im}(A_2)\over\mbox{Re} (A_2)} - 
{\mbox{Im}(A_0)\over \mbox{Re} (A_0)}
 \right] \, .
\eeq
Due to the famous ``$\Delta I=1/2$ rule'', $\eps'/\eps$ is
suppressed by the ratio
%
$\omega = \mbox{Re} (A_2)/\mbox{Re} (A_0) \approx 1/22\, .$
%
The phases of $\eps'$ and $\eps$ turn out to be nearly equal:
%
$\Phi \approx \delta^2_0-\delta^0_0+\frac{\pi}{4}\approx 0 \, .
$
%
The CP--conserving amplitudes $\mbox{Re} (A_I)$, their ratio
$\omega$ and $|\eps|$ are usually set to their experimentally
determined values. A theoretical calculation is then only needed
for $\mbox{Im} (A_I)$.
Using the short--distance Lagrangian \eqn{eq:Leff}, the CP--violating ratio
$\eps'/\eps$ can be written as \cite{munich}
\beq
{\varepsilon^\prime\over\varepsilon} \, = \,
\mbox{Im}\left(V_{ts}^* V_{td}^{\phantom{*}}\right)\, e^{i\Phi}\; 
{G_F\over 2 \vert\varepsilon\vert}\; {\omega\over |\mbox{Re}(A_0)|}\;
\left [P^{(0)}\, (1-\Omega_{IB}) - {1\over \omega} \,P^{(2)}\right ]\, ,
\label{EPS}
\eeq
where the quantities
%
$P^{(I)}= \sum_i\, y_i(\mu)\;
\langle (\pi\pi )_I\vert Q_i\vert K\rangle 
$
%
contain the contributions from hadronic matrix elements 
with isospin $I$,
%
$\Omega_{IB} = (1/ \omega)\,
{\mbox{Im}(A_2)_{\mbox{\small{IB}}}  / \mbox{Im}(A_0)}
$
%
parameterizes isospin breaking corrections and $y_i(\mu)$ are the
 CP--violating parts of the Wilson coefficients: 
$\! C_i(\mu)\! = \! z_i(\mu)+\tau\, y_i(\mu)$ with $\tau=-V_{td}V_{ts}^*/V_{ud}V_{us}^*$.
The factor $1/\omega$ enhances the relative weight of the $I=2$ contributions.
In the Standard Model, $P^{(0)}$ and $P^{(2)}$ turn out to
be dominated  respectively by the contributions from 
the QCD penguin operator $Q_6$ and
the electroweak penguin operator $Q_8$~\cite{trieste},
\beq
\begin{array}{lr}
Q_{6}  =  \left( \overline{s}_{\alpha} d_{\beta}  \right)_{\rm V-A}
   \sum_{q} ( \overline{q}_{\beta}  q_{\alpha} )_{\rm V+ A}
\, , &\quad
Q_{8}  =  \frac{3}{2} \left( \overline{s}_{\alpha}
                                    d_{\beta} \right)_{\rm V-A}
  \sum_{q} e_q \,\left( \overline{q}_{\beta} q_{\alpha}\right)_{\rm V+ A}
\, .
\end{array} 
\label{eq:ope}
\eeq
A recent improved calculation of $\Omega_{IB}^{\pi^0\eta}$
at $\cO(p^4)$ in $\chi$PT has found the result \cite{EMNP:00}
\beq\label{eq:omIB}
\Omega_{IB}^{\pi^0\eta}\, =\, 0.16\pm 0.03 \, .
\eeq
\section{Chiral Perturbation Theory}
Below the resonance region   and using global symmetry considerations one can
define an effective field theory in terms of the QCD Goldstone bosons
$(\pi, K, \eta)$. The $\chi$PT  formulation of the SM~\cite{WE:79,GL:85,EC:95}
describes
the meson--octet dynamics through a perturbative  expansion in powers of the
ratio of
momenta and quark masses over the chiral symmetry breaking scale
$(\Lambda_\chi\sim 1 {\rm GeV})$. The operator content of the theory is fixed
by chiral symmetry.
At lowest order,
the most general effective bosonic weak  Lagrangian, with the same
$SU(3)_L\otimes SU(3)_R$ transformation properties and quantum numbers
as the short--distance Lagrangian \eqn{eq:Leff}, contains three
terms transforming as $(8_L,1_R)$, $(27_L,1_R)$ and  $(8_L,8_R)$ whose
corresponding couplings are denoted by  $g_8$, $g_{27}$ and $g_{ew}$.
%
%


The isospin amplitudes $\cA_I$ have been
computed up to next--to--leading order in the chiral expansion
\cite{KA91}--\cite{EKW:93}. 
Decomposing the isospin amplitudes according  to their
representation components
${ \cA}_I = \Sigma_R { \cA}_I^{(R)}$, 
the results of those calculations can be written in the form
(the expressions for $\cA_0^{(ew)}$,  $\cA_0^{(27)}$, $\cA_2^{(27)}$ can be
found in ref.~\cite{PPS:01}):
\bea 
\label{ONELOOP_8} 
 \cA_0^{(8)}  &=&
-{G_F\over \sqrt{2}} V^{\phantom{\ast}}_{ud}V^\ast_{us} 
\;\sqrt{2}\, f_\pi \; g_8\,
  (M_K^2-M_\pi^2)\,
 \left[1+\Delta_L{\cal A}_0^{(8)} +\Delta_C{\cal A}_0^{(8)}
 \right] \ ,
 \nn \\
 \cA_2^{(ew)} &=&
{G_F\over \sqrt{2}} V^{\phantom{\ast}}_{ud}V^\ast_{us} 
\; {2\over 3}\, 
    e^2\, f^3_\pi \;  g_8 \,\left[
 g_{ew}\,\left( 1 +\Delta_L{\cal A}_{2}^{(ew)}\right)
  +\Delta_C{\cal A}_{2}^{(ew)} 
 \right] \quad .
\eea 
These  formulae contain the
chiral one--loop corrections  $\Delta_L\cA_I^{(R)}$, 
and local contributions $\Delta_C\cA_I^{(R)}$
from ${\cal O}(p^4)$ $\chi$PT counterterms.

It is convenient to rewrite these amplitudes
in the form 
%
${\cal A}_I^{(R)} \; = \; {\cal A}_I^{(R)\infty} \;\times\;\:
{\cal C}_I^{(R)} \, ,
$
%
where ${\cal A}_I^{(R)\infty}$ is the contribution at leading order in the
large--$N_c$ expansion while  the factors ${\cal C}_I^{(R)}$ represent the
next--to--leading order (NLO) correction  in the same expansion.
The chiral loop contributions  are  NLO corrections in $1/N_c$.
In order to determine ${\cal A}_I^{(R)\infty}$ one needs only to match
properly $\chi PT$ with the effective short distance Lagrangian  in
eq.~(\ref{eq:Leff}) and so determine the $\chi PT$ couplings. As an example we
have (a more complete list can be found in ref.~\cite{PPS:01}):
\bea
\nn
\lefteqn{
g_8^\infty\,\left[ 1 + \Delta_C\cA_0^{(8)} \right]^\infty\, =} &&
\nonumber\\ &&
\left\{ 
 -{2\over 5}\,C_1(\mu)+{3\over 5}\,C_2(\mu)+C_4(\mu)
- 16\, L_5\, C_6(\mu)\,
\left[ {M_K^2 \over (m_s + m_q)(\mu)\, f_\pi}\right]^2\right\}\,
f_0^{K\pi}(M_\pi^2)   \, ,\qquad\quad
\\  &&\nonumber\\ \label{eq:NC_results}
\lefteqn{
e^2\, g_8^\infty\, \left[ g_{ew} + \Delta_C\cA_2^{(ew)}\right]^\infty
 \, = \,
-3\, C_8(\mu)\,\left[ {M_K^2 \over (m_s + m_q)(\mu)\, f_\pi}\right]^2
\left[1 + {4 L_5\over f_\pi^2} M_\pi^2\right]
}&&\hfill\mbox{}
\nonumber\\ &&\hspace{4.3cm}\mbox{}
+ {3\over 2} \left[ C_7 - C_9 - C_{10}\right]\!(\mu)\:
{M_K^2-M_\pi^2\over f_\pi^2}\; f_0^{K\pi}(M_\pi^2) \, ,
\eea
where $f_0^{K\pi}(M_\pi^2)\approx 1 + 4 L_5\, M_\pi^2/f_\pi^2$ is the $K\pi$
scalar form factor at the pion mass scale, $L_5$ is a coupling of the strong  ${\cal
  O}(p^4)$ scalar Lagrangian and $m_q\equiv m_u=m_d$.
In the limit $N_c\rightarrow \infty$,
$L_5^\infty=(1/4) f_\pi^2(f_K/f_\pi-1)/(M_K^2-M_\pi^2)\approx 2.1\cdot 10^{-3}$
and $f_0^{K\pi}(M_\pi^2)\approx 1.02$.

These
results are equivalent to the standard large--$N_C$ evaluation of the
usual bag parameters $B_i$.
In particular, for $\eps'/\eps$, where only the imaginary part of the
$g_i$ couplings matter [i.e. Im($C_i$)], the leading order large--$N_c$ estimate
amounts to $B_8^{(3/2)}\approx B_6^{(1/2)}=~1$. Therefore, up to minor
variations of some input parameters, the corresponding $\eps'/\eps$
prediction, obtained at lowest order in both the $1/N_C$ and
$\chi$PT expansions, reproduces the published results of the Munich
\cite{munich} and Rome \cite{rome} groups.
Thus at this order there is a large numerical cancellation between the $I=0$
and $I=2$ contributions, leading to an accidentally small value of $\eps'/\eps$.

Notice that the strong phase shifts are induced by chiral loops and, thus,
they are exactly zero at this leading order approximation.

The large--$N_C$ limit has been only applied to the matching between
the 3--flavor quark theory and $\chi$PT.
The evolution from the electroweak
scale down to $\mu < m_c$ has to be done without any unnecessary expansion
in powers of $1/N_C$; otherwise, one would miss large corrections
of the form ${1\over N_C} \ln{(M/m)}$, with $M\gg m$ two widely
separated scales \cite{BBG87}.
Thus, the Wilson coefficients contain the full $\mu$ dependence.

At large--$N_c$ the operators $Q_i$ ($i\not=6,8$) factorize into products of 
left-- and right--handed vector currents, 
which are renormalization--invariant quantities.
The matrix element of each single current
represents a physical observable which can be directly measured;
its $\chi$PT realization just provides a low--energy expansion in
powers of masses and momenta.
Thus, the large--$N_C$ factorization of these operators
does not generate any scale dependence. Since the anomalous
dimensions of $Q_i$ ($i\not=6,8$) vanish when $N_C\to\infty$ \cite{BBG87},
a very important ingredient is lost in this limit \cite{PI:89}.
To achieve a reliable expansion in powers of $1/N_C$,
one needs to go to the next order where this physics is captured
\cite{PI:89,PR:91}. This is the reason why the study of the $\Delta I=1/2$
rule has proved to be so difficult. Fortunately, these operators
are numerically suppressed in the $\eps'/\eps$ prediction.

The only anomalous dimension components which survive when $N_C\to\infty$
are the ones corresponding to $Q_6$ and $Q_8$ \cite{BBG87,BG:87}.
One can then expect that the matrix elements of these two operators
are well approximated by this limit
 \cite{PI:89,PR:91,JP:94}.
These operators  factorize into color--singlet
scalar and pseudoscalar currents, which are $\mu$ dependent.
This generates the factors $\langle\bar q q\rangle^{(2)}(\mu) \approx
- M_K^2\, f_\pi^2/ (m_s+m_q)(\mu)$  which exactly cancel the $\mu$ dependence of
$C_{6,8}(\mu)$ at large--$N_C$ 
\cite{BBG87,PI:89,PR:91,BG:87,JP:94,dR:89}.
It remains a dependence at next-to-leading order.
While the real part of $g_8$ gets its main contribution
from $C_2$, Im($g_8$) and Im($g_8\,g_{ew}$)
are governed  by $C_6$ and $C_8$, respectively.
Thus, the analyses of the CP--conserving and CP--violating amplitudes
are very different. There are large $1/N_C$ corrections to Re($g_i$)
\cite{PI:89,PR:91,JP:94}, which are needed to understand
the observed enhancement of the $(8_L,1_R)$ coupling.
On the contrary, the large--$N_C$ limit can be expected to give a
good estimate of Im($g_i$).

\section{Chiral loop corrections}

The    large--$N_c$ amplitudes in eq.~(\ref{eq:NC_results}) do not contain any strong phases
$\delta^I_0$.
Those phases originate in the final rescattering of the two pions and,
therefore, are generated by chiral loops which are of higher order in
the $1/N_C$ expansion.
Since the strong phases are quite large, specially in the isospin--zero
case, one should expect large higher--order unitarity corrections.
The multiplicatively correction factors ${\cal C}_I^{(R)}$ contain
the chiral loop contributions we are interested in.
At the one loop, they take the following
numerical values ($
{\cal C}_I^{(R)}\approx  1 +
\Delta_L{\cal A}_I^{(R)}$;
see ref.~\cite{PPS:01} for a
complete list):
\beq
\begin{array}{lcr}
{\cal C}_0^{(8)} = 1.27 \pm 0.05 +  0.46\, i \, , &\quad 
{\cal C}_2^{(27)} =  0.96 \pm 0.05-0.20\, i\, , & \quad
{\cal C}_2^{(ew)} =  0.50  \pm 0.24-0.20\, i\, .
\end{array}
\label{eq:onel}
\eeq
The central values have been evaluated at the chiral renormalization
scale $\nu = M_\rho$. To estimate the corresponding uncertainties
we have allowed the scale $\nu$ to vary between 0.6 and 1~GeV.
The scale dependence is only present in the dispersive contributions
and should cancel with the corresponding $\nu$
dependence of the local $\chi$PT counterterms. However, this dependence is
next-to-leading in $1/N_C$ and, therefore, is not included in our
large--$N_C$ estimate of the $\cO(p^4)$ and $\cO(e^2 p^2)$
chiral couplings.
The $\nu$ dependence of the chiral loops would be
cancelled by the unknown $1/N_C$--suppressed corrections
$\Delta_C{\cal A}_I^{(R)}(\nu)-\Delta_C{\cal A}_I^{(R)\infty}$,
that we are neglecting in the factors ${\cal C}_I^{(R)}$.
The numerical sensitivity of our results to the scale $\nu$
gives then a good estimate of those missing contributions.

The numerical corrections to the 27--plet amplitudes do not have 
much phenomeno\-lo\-gi\-cal interest for CP--violating observables, 
because ${\rm Im}(g_{27}) = 0$.
Remember that the CP--conserving amplitudes ${\rm Re}(A_I)$
are set to their experimentally determined values.
What is relevant for the $\varepsilon'/\varepsilon$
prediction is the 35\% enhancement of the isoscalar octet
amplitude \ Im[$A_0^{(8)}$] \ and the 46\% reduction of \
Im[$A_2^{(ew)}$]. These destroy the
accidental lowest--order cancellation between the $I=0$
and $I=2$ contributions, generating a sizeable enhancement
of $\varepsilon'/\varepsilon$.

A complete $\cO(p^4)$ calculation \cite{EMNP:00,EIMNP:00}
of the isospin--breaking parameter $\Omega_{IB}$ is  not yet
available. The value 0.16 quoted in eq.~(\ref{eq:omIB}) 
only accounts for the contribution from $\pi^0$--$\eta$ 
mixing \cite{EMNP:00}
and should be corrected by the effect of chiral loops. 
Since $|{\cal C}_2^{(27)}| \approx 0.98\pm 0.05$,
one does not expect any large correction of \ ${\rm Im}(A_2)_{IB}$,
while we know that
Im[$A_0^{(8)}$] gets enhanced by a factor 1.35.
Taking this into account, one gets the corrected value
%
$\Omega_{IB} \;\approx\;\Omega_{IB}^{\pi^0\eta}\;
\left|{{\cal C}_2^{(27)}/ {\cal C}_0^{(8)}}\right|
\; = \; 0.12 \pm 0.05 \, ,
$
%
where the quoted error is an educated theoretical guess.
This value agrees with the result 
$\Omega_{IB} = 0.08 \pm 0.05\pm 0.01$,
obtained in ref.~\cite{MW:00} by using three different
models \cite{trieste,EKW:93,PR:91,EPR:91,GV:99,EGPR:89}
to estimate the relevant $\cO(p^4)$ chiral couplings.
\section{FSI at higher orders}
\label{sec:FSI}

Given the large size of the one-loop contributions, one should
worry about higher--order chiral corrections.

The large one-loop FSI correction to the isoscalar amplitudes 
is generated by large infrared
chiral logarithms involving the light pion mass \cite{PP:00}.
These logarithms are universal, i.e. their contribution depends exclusively 
on the quantum numbers of the two pions in the final state \cite{PP:00}. 
As a result, they give the same correction to all isoscalar amplitudes.
Identical logarithmic contributions appear in the scalar pion form factor
\cite{GL:85}, where they completely dominate the $\cO(p^4)$  $\chi$PT
correction.

Using analyticity and unitarity constraints \cite{GP:97}, these
logarithms can be exponentiated to all orders in the chiral
expansion \cite{PP:00}.
The result can be written as:
%
${\cal C}_I^{(R)} \;\equiv\; {\cal C}_I^{(R)}(M_K^2)
\; =\;\Omega_I(M_K^2,s_0) \; {\cal C}_I^{(R)}(s_0)\, .
$
%
The Omn\`es \cite{GP:97,OMNES,TR:88} exponential
\beq\label{eq:omega}
\Omega_I(s,s_0) \;\equiv\;
e^{i\delta^I_0(s)}\; \Re_I(s,s_0) \; =\;
 \exp{\left\{ {(s-s_0)\over\pi}\int
{dz\over (z-s_0)} {\delta^I_0(z)\over (z-s-i\epsilon)}\right\}}
\eeq
provides an evolution of ${\cal C}_I^{(R)}(s)$ from an arbitrary
low--energy point $s_0$ to 
$s\equiv\left(p_{\pi_1}+p_{\pi_2}\right)^2=M_K^2$.
The physical amplitudes are of course independent of the
subtraction point $s_0$.
Intuitively, what the Omn\`es solution does is to correct a local weak
$K\to \pi\pi$ transition with an infinite chain of pion--loop bubbles,
incorporating the strong $\pi\pi\to\pi\pi$ rescattering to all orders
in $\chi$PT.
The Omn\`es exponential only sums a particular type of higher--order
Feynman diagrams, related to FSI. 
Nevertheless, it allows us to perform a reliable estimate 
of higher--order effects because it does sum the most important
corrections.
Moreover, the Omn\`es exponential enforces the decay amplitudes to have
the right physical phases.

The Omn\`es resummation of chiral logarithms is uniquely 
determined up to a polynomial (in $s$) ambiguity
\cite{PP:00,GP:97,BGKO:91},
which has been solved with the large--$N_C$ amplitude
$\cA_I^{(R)\infty}$. The exponential only sums the
elastic rescattering of the final two pions, which is responsible
for the phase shift. Since the kaon mass is smaller than the
inelastic threshold, the virtual loop corrections from other 
intermediate states
($K\to K\pi, K\eta, \eta\eta, K\bar K \to\pi\pi$)
can be safely estimated at the one loop
level; they are included in ${\cal C}_I^{(R)}(s_0)$.

Taking the chiral prediction for $\delta^I_0(z)$ and expanding
$\Omega_I(M_K^2,s_0)$ to $\cO(p^2)$, 
%
%
one should reproduce the one-loop $\chi$PT result.
This determines the factor ${\cal C}_I^{(R)}(s_0)$ to
$\cO(p^4)$ in the chiral expansion.
%
%
It remains a local ambiguity at higher orders
\cite{PP:00,GP:97,BGKO:91}.
To estimate the remaining sensitivity to those higher order corrections,
we have changed the subtraction point between $s_0=0$ and  
$s_0=3 M_\pi^2$ and have included the resulting fluctuations
in the final uncertainties. 
At $\nu=M_\rho$, we get the following
values for the resummed loop corrections ($\left|{\cal C}_I^{(R)}\right|  = 
\Re_I(M_K^2,s_0)\; {\cal C}_I^{(R)}(s_0)$):
\bea
\label{eq:CI_MK}
\left|{\cal C}_0^{(8)}\right|  = 1.31 \pm 0.06\; , &\quad
\left|{\cal C}_2^{(27)}\right|  = 1.05 \pm 0.05\; , &\quad
\left|{\cal C}_2^{(ew)}\right|  =  0.62 \pm 0.05\; .
\eea
These results agree within errors with the one-loop
chiral calculation of the moduli of the isospin amplitudes,
indicating a good convergence of the chiral expansion.

\section{Final results}
\label{sec:numerics}

The infrared effect of chiral loops generates an important enhancement
of the isoscalar $K\to\pi\pi$ amplitude. This effect gets amplified
in the prediction of $\varepsilon'/\varepsilon$, because 
at lowest order (in both $1/N_C$ and the chiral expansion) there
is an accidental numerical cancellation between the $I=0$ and $I=2$
contributions. Since the chiral loop corrections destroy this cancellation,
the final result for $\varepsilon'/\varepsilon$ is dominated by the
isoscalar amplitude. Thus, the Standard Model prediction for
$\varepsilon'/\varepsilon$ is finally governed by the matrix element
of the gluonic penguin operator $Q_6$.

A detailed numerical analysis has been provided in
ref.~\cite{PPS:01}.
The short--distance Wilson coefficients have been evaluated
at the scale $\mu = 1$ GeV. Their associated uncertainties
have been estimated through the sensitivity to 
changes of $\mu$ in the range $M_\rho < \mu < m_c$
and to the choice of $\gamma_5$ scheme.
Since the most important $\alpha_s$ corrections appear at the 
low--energy scale $\mu$, the strong coupling
has been fixed  at the $\tau$ mass,
where it is known~\cite{pich1} with about a few percent level of accuracy:
$\alpha_s(m_\tau)=0.345\pm 0.020$.
The values of $\alpha_s$ at the other needed scales can be
deduced through the standard renormalization group evolution.

Taking the experimental value of $\varepsilon$,
the CP--violating ratio $\varepsilon'/\varepsilon$ 
is proportional to the CKM factor
Im$(V^*_{ts} V^{\phantom{*}}_{td}) = (1.2\pm 0.2) \cdot 10^{-4}$
\cite{anamar}. 
This number is sensitive to the input values of
several non-perturbative hadronic parameters adopted 
in the usual unitarity triangle analysis; thus, it is subject to
large theoretical uncertainties which are difficult to quantify
\cite{PP:95}.
Using instead the theoretical prediction of $\varepsilon$, this CKM
factor drops out from the ratio $\varepsilon'/\varepsilon$; 
the sensitivity
to hadronic inputs is then reduced to the explicit remaining dependence
on the $\Delta S=2$ scale--invariant bag parameter $\hat{B}_K$.
In the large--$N_C$ limit, $\hat{B}_K = 3/4$.
We have performed the two types of numerical analysis, obtaining
consistent results. This allows us to better estimate  the theoretical
uncertainties, since the two analyses have different sensitivity
to hadronic inputs.

The final result quoted in ref.~\cite{PPS:01} is:
\beq
\mbox{\rm Re}\left(\varepsilon'/\varepsilon\right) \; =\;  
\left(1.7\pm 0.2\, {}_{-0.5}^{+0.8} \pm 0.5\right) \cdot 10^{-3}
\; =\; \left(1.7\pm 0.9\right) \cdot 10^{-3}\, .
\label{eq:final_result}
\eeq
The first error comes from the short--distance evaluation
of Wilson coefficients and the choice of the low--energy
matching scale $\mu$.
The uncertainty coming from varying  the strange quark mass 
in the interval $(m_s+ m_q)(1\, \rm{GeV})=156\pm 25\, \rm{MeV}$
\cite{ms2000}
 is indicated 
by the second error.
The most critical step is the matching between the short-- and 
long--distance descriptions. 
We have performed this matching at leading
order in the $1/N_C$ expansion, where the result is known
to $\cO(p^4)$ and $\cO(e^2p^2)$ in $\chi$PT.
This can be expected to provide a good approximation to the matrix 
elements of the leading $Q_6$ and $Q_8$ operators. Since all ultraviolet
and infrared logarithms have been resummed, our educated guess for
the theoretical uncertainty associated with $1/N_C$ corrections
is $\sim 30\%$ (third error).

A better determination of the strange quark mass would allow
to reduce the uncertainty to the 30\% level.
In order to get a more accurate prediction, it would be necessary to have
a good analysis of next--to--leading $1/N_C$ corrections. This is
a very difficult task, but progress in this direction can be
expected in the next few years 
\cite{trieste,BP:00,PR:91,DG:00,NLO_NC,latka}.

To summarize, using a well defined computational scheme,
it has been possible to pin down the value of 
$\varepsilon'/\varepsilon$ with an acceptable accuracy.
Within the present uncertainties, the resulting
Standard Model theoretical prediction \eqn{eq:SMpred}
is in good agreement with the measured experimental
value \eqn{eq:exp}.

I.S. wishes to thank the organizers of EPS2001 for the nice meeting.
This work has been partially supported by the TMR Network ``EURODAPHNE''
(Contr.No. ERBFMX--CT98--0169) and by DGESIC, Spain (Grant No. PB97--1261).



\begin{thebibliography}{99}
\bibitem{PPS:01}E.~Pallante, A.~Pich, I.~Scimemi, Nucl. Phys. B in press, hep--ph/0105011.
\bibitem{PP:00} E. Pallante and A. Pich,  Phys. Rev. Lett.
    84 (2000) 2568,    
 Nucl. Phys. B592 (2000) 294.



\bibitem{na48}
  
  NA48 collaboration (V. Fanti {\em et al.}), Phys. Lett. B465
  (1999) 335,  
  (A. Lai {\em et al.}) hep--ex/0110019.

\bibitem{ktev} KTeV collaboration (A. Alavi--Harati {\em et al.}) 
               \prl  83 (1999) 22; R. Kessler, proceedings of Lepton-Photon
               2001, hep--ex/0110020.
  

\bibitem{NA31} NA31 collaboration
  (H.~Burkhardt {\em et al.}), Phys. Lett. {B206} (1988) 169;
  (G.D.~Barr {\em et al.}), Phys. Lett. {B317} (1993) 233.

\bibitem{E731} E731 collaboration (L.K. Gibbons {\em et al.}), Phys. Rev. Lett.
  {70} (1993) 1203.

\bibitem{munich}
 A.J. Buras {\em et al.}, Nucl. Phys. B592 (2001) 55;  
 S.~Bosch {\em et al.}, Nucl. Phys. B565 (2000) 3;   
 G.~Buchalla {\em et al.}, Rev. Mod. Phys. 68 (1996) 1125;
 A.J.~Buras {\em et al.}, Nucl. Phys. B408 (1993) 209,
 Phys. Lett. B389 (1996) 749.
 

\bibitem{rome}
  M. Ciuchini {\em et al.}, Nucl. Phys. B (Proc. Suppl.) 99 (2001) 27; 
  hep-ph/9910237;
  Nucl. Phys. B523 (1998) 501, B415 (1994) 403;
  Phys. Lett. B301 (1993) 263.

\bibitem{trieste}
 S.~Bertolini {\em et al.}, Rev. Mod. Phys. 72 (2000) 65; 
 Phys. Rev. D63 (2001) 056009;   
 Nucl. Phys. B449 (1995) 197, B476 (1996) 225, B514 (1998) 63, 93;
  M.~Fabbrichesi, Phys. Rev. D62 (2000) 097902.

\bibitem{dortmund}
 T. Hambye {\em et al.},  hep-ph/0001088;    
 Nucl. Phys. B564 (2000) 391;
 Eur. Phys. J. C10 (1999) 271;
 Phys. Rev. D58 (1998) 014017;
 Y.-L. Wu, Phys. Rev. D64 (2001) 016001.  

\bibitem{BP:00} J.~Bijnens and J.~Prades, JHEP 06 (2000) 035;
  Nucl. Phys. (Proc. Suppl.) 96 (2001) 354; 
J.~Bijnens, E.~Gamiz and J.~Prades, JHEP 10 (2001) 009.

\bibitem{dubna} A.A. Belkov {\em et al.}, hep-ph/9907335;
 H.Y. Chen, Chin. J. Phys. 38 (2000) 1044;  
S. Narison, Nucl. Phys. B593 (2001) 3. 







\bibitem{WI:69} K.G. Wilson Phys.~Rev.  179 (1969) 1499; 
  W. Zimmermann, {\it Lectures on Elementary Particles and Quantum
  Field Theory}, Brandeis Summer Institute (1970) Vol.~1 (MIT Press,
  Cambridge, MA, 1970).
  
\bibitem{RGroup}
  E.C.G. Stueckelberg and A. Peterman, Helv. Phys. Acta 26 (1953) 499;
  M. Gell-Mann and F.E. Low, Phys. Rev. 95 (1954) 1300;
  C. Callan Jr., Phys. Rev. D2 (1970) 1541;
  K.~Symanzik, Commun. Math. Phys. 18 (1970) 227; 23 (1971) 49.






\bibitem{buras1} 
  A.J.~Buras {\em et al.}, Nucl. Phys. B400 (1993) 37,
  75, 
  B370 (1992) 69.


\bibitem{ciuc1} M.~Ciuchini {\em et al.}, \pl { B301} (1993) 263;
    Z. Phys. C68 (1995) 239.

\bibitem{GM:91} J. Gasser, U-G. Meissner, Phys. Lett. B258 (1991) 219. 



\bibitem{EMNP:00} G. Ecker, G. M\"uller, H. Neufeld and A. Pich,
    Phys. Lett. B477 (2000) 88.


\bibitem{WE:79} S. Weinberg, Physica 96A (1979) 327.

\bibitem{GL:85}
 J. Gasser and H. Leutwyler, Ann. Phys., NY 158 (1984) 142;
     Nucl. Phys. B250 (1985) 456; 517; 539.

\bibitem{EC:95}
 G. Ecker, Prog. Part. Nucl. Phys. 35 (1995) 1;
 A. Pich, Rep. Prog. Phys. 58 (1995) 563;
 U.-G. Meissner, Rep. Prog. Phys. 56 (1993) 903.





\bibitem{KA91}
  J. Kambor, J. Missimer and D. Wyler, Nucl. Phys. B346 (1990) 17;
  Phys. Lett. B261 (1991) 496; 
  J. Kambor {\em et al.}, Phys. Rev. Lett. 68 (1992) 1818.

\bibitem{BPP}
  J. Bijnens, E. Pallante and J.~Prades, Nucl. Phys. B521 (1998) 305;
  E. Pallante, JHEP 01 (1999) 012.



\bibitem{EIMNP:00} G. Ecker, G. Isidori, G. M\"uller, H. Neufeld
   and A. Pich, Nucl. Phys. B591 (2000) 419;  
   work in progress.  


\bibitem{CDG:99} V. Cirigliano, J.F. Donoghue and E. Golowich,
   Phys. Lett. B450 (1999) 241; Phys. Rev. D61 (2000) 093001, 093002.

\bibitem{CG:00} V. Cirigliano and E. Golowich, Phys. Lett. B475 (2000) 351.

\bibitem{EKW:93} G. Ecker, J. Kambor and D. Wyler, Nucl. Phys. B394
   (1993) 101.

\bibitem{BBG87} W.A. Bardeen, A.J. Buras and J.-M. G\'erard,
  Nucl. Phys. B293 (1987) 787; Phys. Lett. B211 (1988) 343,
  B192 (1987) 138, B180 (1986) 133;
  A.J. Buras and J.-M.~G\'erard, Nucl. Phys. B264 (1986) 371.

\bibitem{PI:89} A. Pich, Nucl. Phys. B (Proc. Suppl.) 7A (1989) 194.

\bibitem{PR:91} A. Pich and E. de Rafael, Nucl. Phys. B358 (1991) 311;
  Phys. Lett. 374 (1996) 186.

\bibitem{BG:87}
  A.J.~Buras and J.-M. G\'erard, Phys. Lett. B192 (1987) 156.

\bibitem{JP:94} M. Jamin and A. Pich, Nucl. Phys. B425 (1994) 15.


\bibitem{dR:89} E. de Rafael, Nucl. Phys. B (Proc. Suppl.) 7A (1989) 1.

\bibitem{MW:00} K. Maltman and C.E. Wolfe, Phys. Lett. B482 (2000) 77;
  Phys. Rev. D63 (2001) 014008.  

\bibitem{EPR:91}
  G. Ecker, A. Pich and E. de Rafael, Phys. Lett. B237 (1990) 481.

\bibitem{GV:99} S. Gardner and G. Valencia, Phys. Lett. B466 (1999) 355.



\bibitem{EGPR:89} 
 G. Ecker, J. Gasser, A. Pich and E. de Rafael, Nucl. Phys. B321 (1989) 
 311; G. Ecker, J.~Gasser, H. Leutwyler, A. Pich and E. de Rafael, 
 Phys. Lett. B223 (1989) 425.


\bibitem{GP:97} F. Guerrero and A. Pich, Phys. Lett. B412 (1997) 382.

\bibitem{OMNES} R. Omn\`es, {\em Nuovo Cimento} 8 (1958);
 N.I. Muskhelishvili, {\it Singular Integral Equations},
  Noordhoof, Groningen, 1953.

\bibitem{TR:88} T.N. Truong, Phys. Lett. B207 (1988) 495.

\bibitem{BGKO:91} M. B\"uchler {\em et al.}, Phys. Lett. B521 (2001) 22, 
   29.


\bibitem{pich1} A. Pich, {\it Tau Physics: Theoretical Perspective},
  Proc. 6th Int. Workshop on Tau Lepton Physics (Victoria, Canada, 2000),
  Nucl. Phys. B (Proc. Suppl.) 98 (2001) 385;  
  {\it Tau Physics}, Proc. XIX International
 Symposium on Lepton and Photon Interactions at High Energies (Stanford, 1999),
 eds. J.~Jaros and M.~Peskin (World Scientific, Singapore, 2000) 157
 [hep-ph/9912294].


\bibitem{anamar}
  M. Ciuchini {\em et al.}, Nucl. Phys. B573 (2000) 201, hep-ph/0012308;
  S. Mele, Phys.~Rev.  D59 (1999) 113011;
  F. Parodi, P. Roudeau and A. Stocchi, Nuovo Cim. A112 (1999) 833;  
D.~Atwood, A. Soni,  Phys. Lett. B508 (2001) 17; 
A. H\"ocker, H. Lacker, S. Laplace, F. Le Diberder, 
Eur. Phys. J. C21 (2001) 225.

\bibitem{PP:95} J. Prades and A.~Pich, Phys. Lett. B346 (1995) 342; 
  J.L.~Rosner, Nucl. Instrum. Meth. A462 (2001) 304;  
M.~Fabbrichesi, Phys. Rev. D62 (2000) 097902. 

\bibitem{ms2000} 
M. Jamin, J.A. Oller and A. Pich, hep--ph/0110194;
S. Chen, M. Davier, E. G\'amiz, A.~H\"ocker, A. Pich and J. Prades,
(Eur. Phys. J. C in press)
hep-ph/0105253;
Proc. 6th Int. Workshop on Tau Lepton Physics (Victoria, 
Canada, 2000), Nucl. Phys. B (Proc. Suppl.) 98 (2001) 319;
A. Pich and J.~Prades, JHEP 10 (1999) 004, JHEP 06 (1998) 013;
ALEPH Collab., Eur. Phys. J. C11 (1999) 599, C10 (1999) 1;
J. Kambor and K. Maltman, Phys. Rev. D62 (2000) 093023;
K.G. Chetyrkin {\em et al.}, Nucl. Phys. B533 (1998) 473;
  M. Jamin, Nucl. Phys. B (Proc. Suppl.) 64 (1998) 250;
  K.G. Chetyrkin {\em et al.}, Phys. Lett. B404 (1997) 337;
  P.~Colangelo {\em et al.}, Phys. Lett. B408 (1997) 340;
  C.A.~Dominguez {\em et al.}, Nucl. Phys. B (Proc. Suppl.) 74 (1999) 313;
  K. Maltman, Phys. Lett. B462 (1999) 195;
  S. Narison, Phys. Lett. B466 (1999) 345;  
   V. Lubicz, Nucl. Phys. B (Proc. Suppl.) 94 (2001) 116;
   R. Gupta and K.~Maltman, hep-ph/0101132;
  J.~Prades, Nucl. Phys. B (Proc. Suppl.) 64 (1998) 253;
  J.~Bijnens, J.~Prades and E.~de Rafael, Phys. Lett. B348 (1995) 226.



\bibitem{DG:00} J.F. Donoghue and E. Golowich, Phys. Lett. B478 (2000) 172;
  V. Cirigliano {\em et al.}, hep--ph/0109113.








\bibitem{NLO_NC} M. Knecht, S. Peris and E. de Rafael, 
Phys. Lett. B508 (2001) 117,    
B457 (1999) 227, B443 (1998) 255;
Nucl. Phys. B (Proc. Suppl.) B86 (2000) 279. 

\bibitem{latka} 
L. Lellouch and M. L\"uscher, Commun. Math. Phys. 219 (2001) 31;
M.F.L. Golterman and E.~Pallante, JHEP 08 (2000) 023;
C.-J.D. Lin {\em et al.}, hep-lat/0104006.




\end{thebibliography}
\end{document}